\documentclass[9pt,twocolumn,twoside]{optica}
\usepackage{physics}
\usepackage{bm}
\setboolean{shortarticle}{false}
\setboolean{minireview}{false}

\title{Spin-orbit coupling in photonic graphene}

\author[1]{Zhaoyang Zhang}
\author[1]{Shun Liang}
\author[1,*]{Feng Li}
\author[1]{Shaohuan Ning}
\author[1]{Yiming Li}
\author[2]{G. Malpuech}
\author[1]{Yanpeng Zhang}
\author[4,5]{Min Xiao}
\author[2,3,*]{D. Solnyshkov}

\affil[1]{Key Laboratory for Physical Electronics and Devices of the Ministry of Education \& Shaanxi Key Lab of Information Photonic Technique, School of Electronic Science and Engineering, Faculty of  Electronic and Information Engineering, Xi’an Jiaotong University, Xi’an 710049, China}
\affil[2]{Institut Pascal, PHOTON-N2, Universit\'e Clermont Auvergne, CNRS, SIGMA Clermont, F-63000 Clermont-Ferrand, France}
\affil[3]{Institut Universitaire de France (IUF), F-75231 Paris, France}
\affil[4]{Department of Physics, University of Arkansas, Fayetteville, Arkansas, 72701, USA}
\affil[5]{National Laboratory of Solid State Microstructures and School of Physics, Nanjing University, Nanjing 210093, China}
\affil[*]{Corresponding authors: felix831204@xjtu.edu.cn, dmitry.solnyshkov@uca.fr}

\begin{abstract}
    We generate experimentally a honeycomb refractive index pattern in an atomic vapor cell using electromagnetically-induced transparency. We study experimentally and theoretically the propagation of polarized light beams in such "photonic graphene".We demonstrate that an effective spin-orbit coupling appears as a correction to the paraxial beam equations because of the strong spatial gradients of the permittivity. It leads to the coupling of spin and angular momentum at the Dirac points of the graphene lattice. Our results suggest that the polarization degree plays an important role in many configurations where it has been previously neglected. 
\end{abstract}

\setboolean{displaycopyright}{true}

\begin{document}

\maketitle

\section{Introduction}
Topological photonics \cite{lu2014topological,ozawa2019topological} is a rapidly growing field, combining fundamental physics and applied optics. The research in this field has brought us new understanding of the fundamental topological properties of optical systems, which are due to the photonic spin-orbit coupling present in various kinds of inhomogeneous photonic systems \cite{Onoda2004,Kavokin2005,bliokh2015quantum}. Photonic spin-orbit coupling (SOC) is a crucial ingredient for solving long-standing problems like optical isolation at a microscopic scale \cite{wang2009observation,Nalitov2015,solnyshkov2018topological,klembt2018exciton,Karki2019} required for the functioning of lasers, opening a new field of topological lasers \cite{Solnyshkov2016,StJean2017,Bahari2017,Bandres2018}.

Photonic graphene is a system of a particular interest. The graphene lattice, studied for more than a half-century \cite{wallace1947band}, was one of the first to demonstrate the striking manifestations of the Dirac physics  \cite{Semenoff1984} such as the Klein tunneling \cite{katsnelson2006chiral}, with enormous potential for applications, which have already found their way to the market \cite{minhyun2019lithium}. In photonics, the Dirac points of graphene-like lattices offer extended possibilities for the manipulation of optical angular momentum \cite{Allen1992} and for the studies of singular optical beams \cite{Rozas1997}. Recent works address various problems, such as beam conversion \cite{Zhang2019}, intervalley scattering \cite{Song2019}, and valley pseudospin dynamics \cite{Song2015}. Different implementations of photonic graphene include coupled waveguides \cite{plotnik2014observation}, microwave resonators \cite{Bellec2013}, photorefractive nonlinear crystals \cite{fleischer2003observation,Song2015}, microcavities \cite{Jacqmin2014,milicevic2015edge,Milicevic2018,klembt2018exciton}, and atomic vapor cells \cite{Zhang2019}, as in the present work. While the main feature of the graphene lattice (the presence of the Dirac cones) is present in all these implementations, other properties can be different. In particular, the SOC in 2D photonic systems such as microcavities is induced by the splitting between TE-TM polarized modes \cite{Kavokin2005}. It is known  to modify the dispersion at the Dirac point \cite{Nalitov2015,Nalitov2015b}, leading to trigonal warping, like in bilayer graphene \cite{novoselov2006unconventional}. In coupled waveguides arrays, usually only a single polarization mode is used and the other polarization can be neglected. In nonlinear crystals and atomic vapor cells, the effects of the SOC on the photonic graphene have not been studied so far.

The evolution of a photonic beam in a spatially varying medium is a particularly important fundamental and applied problem. It is often described in the paraxial approximation of the Helmholtz equation \cite{Shen2003}, especially in the field of nonlinear optics, where it allows to determine the spatial mode profiles. The coupling of polarizations can arise either due to the anisotropy of the material or to its inhomogeneity \cite{wagner1968large}. The former usually couples circular polarizations \cite{Brasselet2009} and was already shown to lead to angular momentum transfer  \cite{Ciattoni2003,fadeyeva2010extreme}, while the latter has not been fully studied so far. In many cases, the intrinsic coupling of polarizations is simply neglected in the paraxial approximation \cite{wagner1968large}. Taking it into account in the calculations of the beam trajectory and properties often leads to spectacular effects, such as the spin Hall effect of light \cite{Onoda2004,Bliokh2009,Aiello2009}. 

The behavior of the polarization of light has been described in the limit of geometric optics in the works of Rytov \cite{Rytov1938}. The Rytov's matrix allows to predict the rotation of the linear polarization. For rays belonging to the class of planar curves (that is, lying in a plane), such rotation is absent: the transverse electric field keeps its polarization in the plane. However, if the ray trajectory becomes three-dimensional (e.g. helix trajectory), the linear polarization starts to rotate. This rotation was linked with anholonomic effects a long time ago \cite{Vladimirskii1941} and was shown to lead to the accumulation of the Berry phase \cite{berry1984quantal} shortly after its discovery \cite{Lipson1990}. However, the corresponding theory was limited to ray tracing, equivalent to considering a point-like particle (beam center of mass) instead of a wave packet (beam envelope), whereas modern research subjects, such as the photonic graphene, clearly require a complete wave theory for the transverse beam evolution. The first attempts to develop such theory have shown the necessity for SOC in the paraxial approximation, but did not lead to a self-consistent system of equations \cite{jisha2017paraxial}.

In this work, we introduce the SOC terms into the paraxial equations for the two transverse polarizations. We experimentally demonstrate that these terms play a particular role in photonic graphene, where they couple the spin and angular momentum at the Dirac points, modifying the angular momentum of the probe beam depending on its polarization.

\section{The model}

The Helmholtz equation for the electric field of an electromagnetic wave in a dielectric medium reads
\begin{equation}
    \nabla^2\bm{E}+k_0^2n^2\bm{E}=0
    \label{Helm}
\end{equation}
In a homogeneous system, this equation does not contain any SOC terms, and the polarizations are decoupled. This allows writing a paraxial equation for a scalar amplitude, corresponding to a single chosen polarization of light (which is conserved). It is very well known that this paraxial equation is equivalent to the time-dependent Schrodinger equation for a wave function of a scalar particle.

However, when the spatial gradients are not negligibly small and when the polarization effects are explicitly studied, the SOC has to be taken into account. It is associated with a coupling of the two polarizations, transverse-electric and transverse-magnetic \cite{Landau8}, which become well-defined in the presence of any gradient (to which they are transverse in addition to being perpendicular to the propagation direction). In geometric optics, the SOC leads to the evolution of the transverse linear polarization along a curved beam \cite{Rytov1938}. This adiabatic evolution has been shown to lead to  dramatic effects, such as the spin Hall effect of light \cite{Onoda2004,Bliokh2008,Bliokh2009}. Coming back to the analogy with quantum mechanics, the geometric optics corresponds to studying a classical particle, whereas the paraxial equation corresponds to considering an equivalent quantum wave packet. Starting from the  evolution of the polarization in the spin Hall effect of light for a beam in the geometric optics limit, that is, for a classical propagating particle, we show how this term is introduced into paraxial equations for the two projections of the electric field amplitude, leading to an original type of SOC. This is especially important for the study of the propagation of beams in such systems as photonic graphene, whose energy bands appear from the quantum-mechanical description.

\begin{figure}[tbp]
\centering
\fbox{\includegraphics[width=\linewidth]{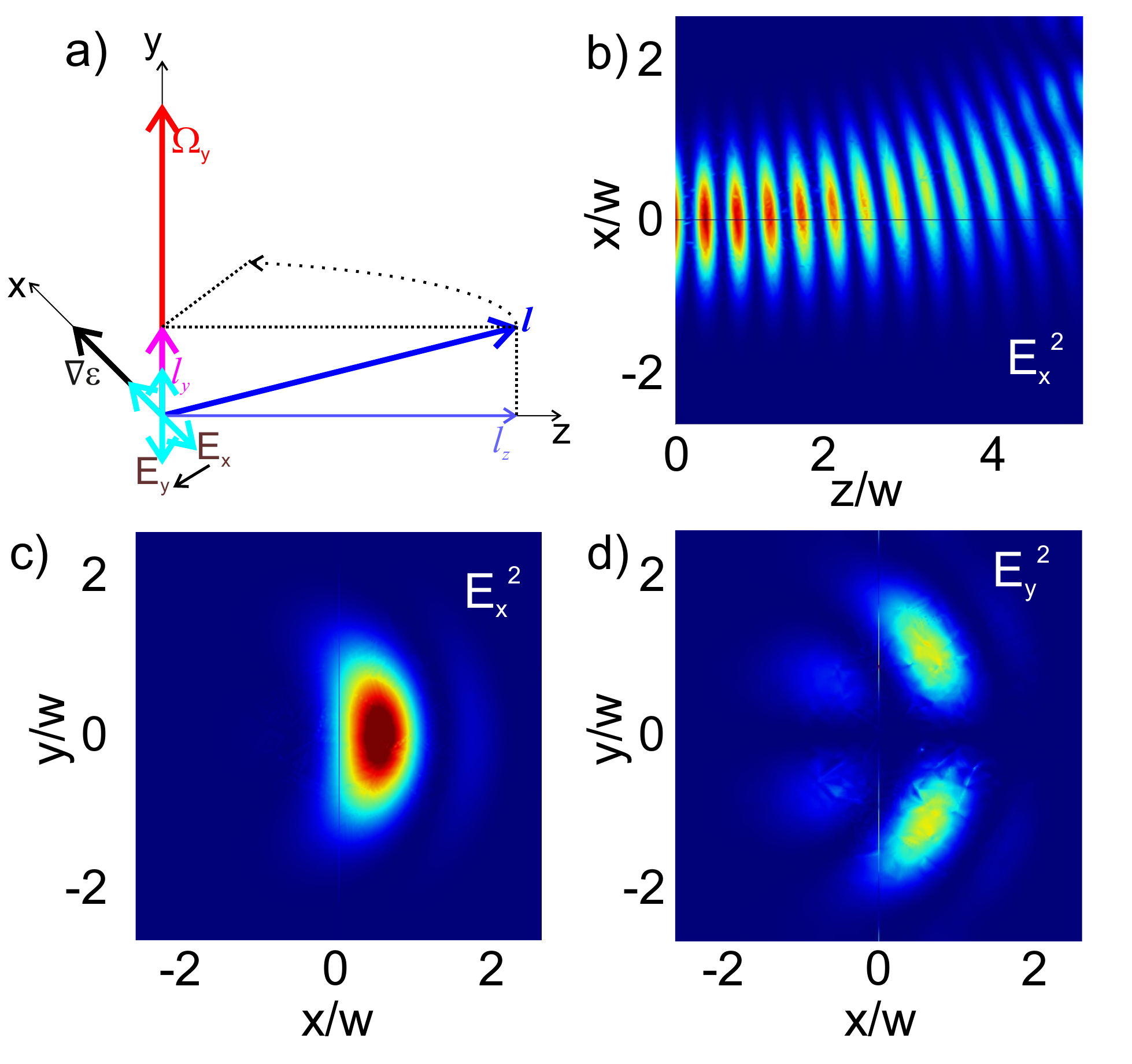}}
\caption{Polarization conversion under a constant gradient. a) Scheme of the effect ($\bm{k}$ is the wave vector of light (direction $\bm{l}$), $\bm{\Omega}$ is the rotation frequency due to the permittivity gradient $\nabla\epsilon$; b) Beam trajectory in the $XZ$ plane (curved due to the gradient $\nabla\epsilon$). Transverse beam profiles: c) in the $XY$ plane in the $E_x$ polarization. Shift along $x$ is due to $\nabla\epsilon$; d) in the $XY$ plane in the $E_y$ polarization entirely generated due to SOC, proportional to $\bm{\Omega}\cdot\bm{k}\sim k_y$ (zero at $y=0$).}
\label{fig0}
\end{figure}

As noted already by Rytov \cite{Rytov1938}, the linear polarization can rotate only for a non-planar beam trajectory. Noting the beam direction as $\bm{l}=\bm{k}/k$ and the vector of its rotation as $\bm{\Omega}$, the relevant quantity describing the part of its rotation which is not in a plane can be written as $\bm{\Omega}\cdot\bm{l}$. This is illustrated by a scheme in Fig.~\ref{fig0}(a). It was shown that the rotation of the transverse polarization adds an extra term to the Helmholtz equation for the transverse electric field
\begin{equation}
    \nabla^2\bm{E}_{\perp}+k_0^2n^2\bm{E}_{\perp}+2ink_0\left(\bm{\Omega}\cdot\bm{l}\right)\left[\bm{l}\times\bm{E}_{\perp}\right]=0
\end{equation}
This term introduced in \cite{Lipson1990} can be understood as an analog of the Coriolis force, appearing in a non-inertial frame, whose non-inertial nature is due precisely to the rotation $\bm{\Omega}$. The geometric optics limit of this equation has allowed to describe the spin Hall effect of light \cite{Bliokh2008}.

In order to include this term into the paraxial equations, we need to link $\bm{\Omega}$ with the transverse field $\bm{E}$ and the dielectric permittivity $\epsilon$ or the dielectric susceptibility $\chi$. Let us consider a beam with its main propagation direction along $z$, containing non-zero $x$ and $y$ wave vector harmonics because of its transverse profile, in presence of a gradient along $x$: $\partial\epsilon/\partial x\neq 0$, which leads to the deviation of the beam from its initial direction: $\Omega_y=  (2n)^{-1}\partial\epsilon/\partial x$. The projection of the rotation frequency on the direction of a particular harmonic  is given by
\begin{equation}
    \bm{\Omega}\cdot\bm{l}=\frac{k_y}{2nk_0}\frac{\partial \epsilon}{\partial x}
    \label{om}
\end{equation}
 In order confirm that the polarization conversion indeed takes place in such conditions, we perform a FDTD numerical simulation of the propagation of a Gaussian beam using COMSOL in the simplest system described above. The results are shown in Fig.~\ref{fig0}(b)-(d). The trajectory of the beam, curved by the gradient $\partial \epsilon/\partial x$, is shown in Fig.~\ref{fig0}(b). The beam is initially excited with $E_x$ polarization only. The transverse profile in this polarization is shown in Fig.~\ref{fig0}(c). The beam is shifted towards positive $x$, as in panel (b). Its shape also changes. But the most important effect is shown by Fig.~\ref{fig0}(d), presenting the cross-polarization $E_y$, appearing only because of the SOC. This requires a non-zero projection of the wave vector $\bm{k}$ on its rotation frequency $\bm{\Omega}$, according to \eqref{om}, which is fulfilled thanks to the presence of non-zero $k_y$ in the narrow initial Gaussian beam. The intensity of converted polarization is linear in $\nabla\epsilon$, confirming the first-order nature of the effect. We note that since the converted signal is proportional to $k_y$, it changes sign at $y=0$: the symmetry of the state is inverted (from symmetric to anti-symmetric). This will be important for the understanding of SOC effects in a periodically modulated medium (lattice).

Generalizing this result to arbitrary gradient directions and substituting $k_{(x,y)}=-i\partial/\partial (x,y)$, one obtains the following SOC term:
\begin{equation}
    \frac{1}{2nk_0}\left(\frac{\partial \epsilon}{\partial x}\frac{\partial E_x}{\partial y}+\frac{\partial \epsilon}{\partial y}\frac{\partial E_y}{\partial x}\right)\bm{l}\times\bm{E}_{\perp}
\end{equation}
As compared with TE-TM SOC in planar cavities \cite{Kavokin2005,CRAS2016} and photonic crystal slabs, the double spatial derivative of the electric field is replaced by a product of the first derivatives of the permittivity and the electric field components.

In atomic systems, the permittivity can be varied through the effect of electromagnetically induced transparency (EIT) \cite{Banacloche1995}.  Another polarization effect which obviously needs to be taken into account in presence of EIT under polarized pumping is the polarization-dependent complex permittivity. This is naturally included in the paraxial equations via $\epsilon_{x,y}$ or susceptibility $\chi_{x,y}$  with real and imaginary parts (marked by $'$ and $''$ correspondingly). The final paraxial equations taking into account the SOC and the polarization-dependent propagation in the linear polarization basis read:
\begin{eqnarray}
\label{paraxpol}
i\frac{\partial E_x}{\partial z}&=&-\frac{1}{2 k_0}\left(\frac{\partial^2}{\partial x^2}+\frac{\partial^2}{\partial y^2}\right)E_x-\frac{k_0\chi_x}{2} E_x\nonumber\\
&+&\frac{1}{2k_0}\left(\frac{\partial \chi'_x}{\partial x}\frac{\partial E_x}{\partial y}+\frac{\partial \chi'_y}{\partial y}\frac{\partial E_y}{\partial x}\right)E_y\\
i\frac{\partial E_y}{\partial z}&=&-\frac{1}{2 k_0}\left(\frac{\partial^2}{\partial x^2}+\frac{\partial^2}{\partial y^2}\right)E_y-\frac{k_0\chi_y}{2} E_y\nonumber\\
&+&\frac{1}{2k_0}\left(\frac{\partial \chi'_x}{\partial x}\frac{\partial E_x}{\partial y}+\frac{\partial \chi'_y}{\partial y}\frac{\partial E_y}{\partial x}\right)E_x
\end{eqnarray}

Of course, the importance of the corrective SOC term depends on the conditions of a given experiment. We will compare the predictions of these paraxial equations with the solution of the full system of Maxwell's equations and with the experimental measurements for a particular system of photonic graphene, and show that, because of the polarization-dependent decay rate, the contribution of the SOC actually becomes dominant.

\section{Experimental implementation of photonic graphene}

\begin{figure}[tbp]
\centering
\fbox{\includegraphics[width=\linewidth]{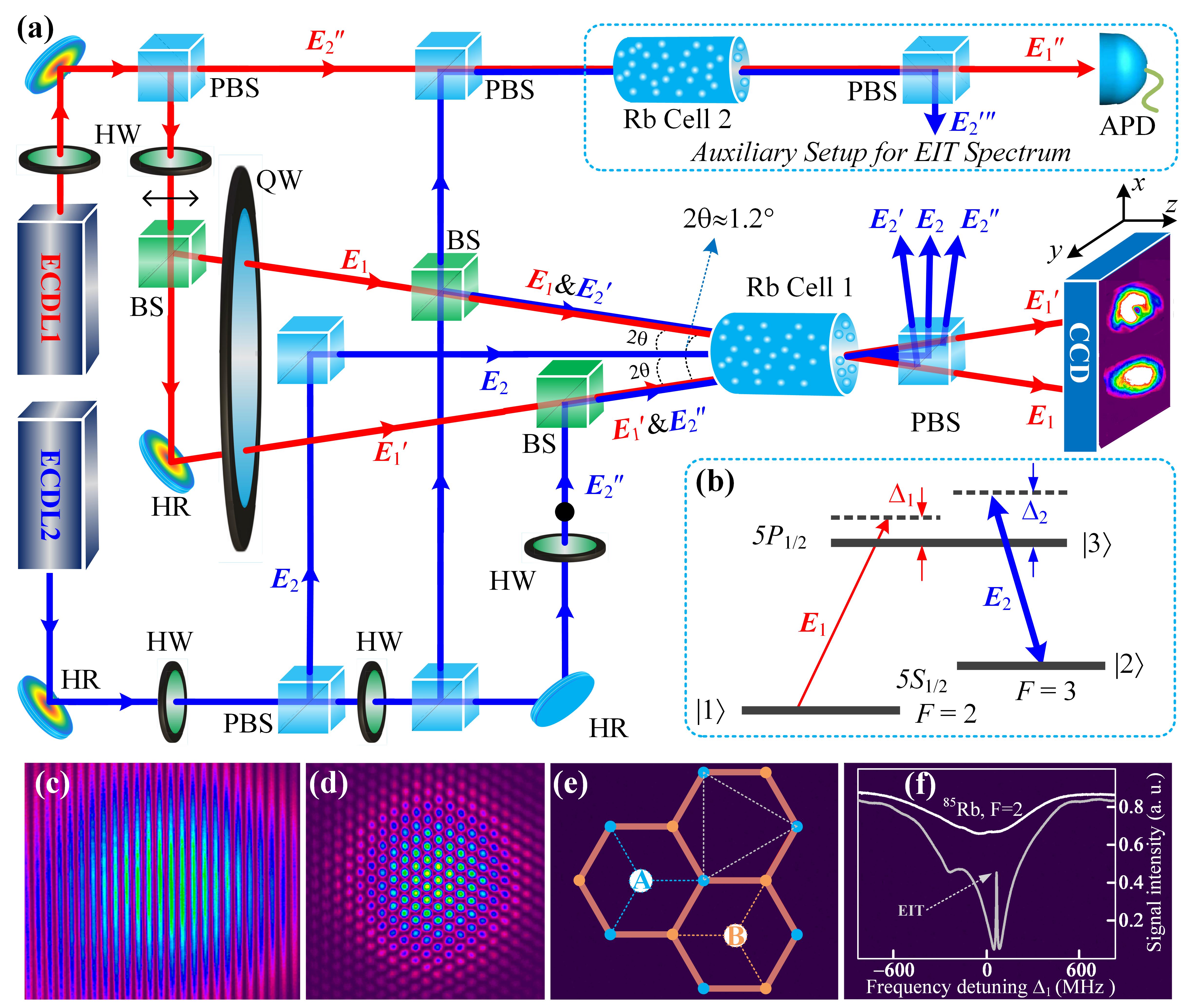}}
\caption{Illustration of the experimental setup and principles. (a) Experimental setup. ECDL: external cavity diode laser; HW: half-wave plate; QW: quarter-wave plate; HR: high-reflectivity mirror; PBS: polarization beam splitter; BS: beam splitter; APD: avalanche photodiode detector; CCD: charge coupled device camera. The beam intensities are controlled by corresponding tunable attenuators. Double-headed arrows and filled dots denote horizontal polarization and vertical polarization of the beams, respectively. (b) The three-level $\Lambda$-type  $^{85}$Rb atomic energy-level structure. Term $\Delta_1$ ($\Delta_2$) is the frequency detuning between the atomic resonance $\ket{1}\to\ket{3}$ ($\ket{2}\to\ket{3}$) and the probe (coupling) frequency. (c) The standing-wave probe field formed by $E_1$ and $E'_1$. (d) The formed hexagonal optical lattice by $E_2$, $E'_2$, and $E''_2$. (e) The schematic diagram for the A- and B-type sublattices on a graphene structure. (f) The observed absorption (upper curve) and EIT (lower curve) spectra from the auxiliary setup [marked by the dashed box in the upper right corner of (a)]. The EIT spectrum is generated by injecting beams $E'_1$ and $E''_2$ (from the same laser as $E_1$ and $E_2$, respectively) into the Rb cell 2 and received by an APD.}
\label{fig1}
\end{figure}

Whilst the SOC plays a significant role in a large variety of optical systems, we investigate its effect in a particularly interesting process: the vortex generation at the Dirac points in photonic graphene. Our experiments are performed in atomic vapors in the regime of EIT. Figure \ref{fig1} shows the scheme of the experiment for exciting a given valley in the photonic graphene lattice. Three Gaussian coupling beams $E_2$, $E'_2$ and $E''_2$ (wavelength $\lambda_2\approx794.97$ nm, frequency $\omega_2$, vertical polarization, Rabi frequencies $\Omega_2$, $\Omega'_2$ and $\Omega''_2$, respectively) from the same external cavity diode laser (EDCL2) symmetrically propagate along the $z$ direction and intersect at the center of the vapor cell with the same angle of $2\theta\approx1.2^\circ$ between any two of them to form a hexagonal optical lattice [see Fig.~\ref{fig1}(d)], acting as the coupling field with an intensity of $|\Omega_c|^2$. This lattice appears as a result of interference, and thus does not suffer from any broadening due to diffraction. The 5~cm long atomic vapor cell wrapped with $\mu$-metal sheets is heated to $80^\circ$ by a heat tape. The co-propagating probe field and coupling field interact with an $\Lambda$-type three-level $^{85}$Rb atomic system [see Fig.~\ref{fig1}(b)], which consists of two hyperfine states $F=2$ (level $\ket{1}$) and $F=3$ ($\ket{2}$) of the ground state $5S_{1/2}$, and an excited state $5P_{1/2}$ ($\ket{3}$). Here, the probe field [Fig.~\ref{fig1}(c)] is established by the interference of two probe beams $E_1$ and $E'_1$ ($\lambda_1\approx794.98$~nm, $\omega_1$, $\Omega_1$ and $\Omega'_1$, respectively) derived from the same ECDL1. A polarization beam splitter (PBS) is mounted in front of the CCD camera to filter out the coupling beam which is set as $y$-polarized throughout the experiment, allowing only the detection of the $x$-polarized component of the probe beam [see Fig.~\ref{fig1}(a)].

With the two-photon resonant condition $\Delta_1-\Delta_2=0$ satisfied, an EIT window \cite{Banacloche1995} can occur on the transmitted probe spectrum of the probe field. For an EIT configuration, the susceptibility $\chi=\chi'+i\chi''$ experienced by the probe field is $\chi\sim|\Omega_c|^{-2}$ \cite{Zhang2018}, and the resulted susceptibility (both real part $\chi'$ and imaginary part $\chi''$ ) exhibits a honeycomb profile with a lattice constant $a\approx 25$~$\mu$m, which corresponds to an inverted hexagonal $|\Omega_c|^2$ \cite{Zhang2019}. For a three-level EIT atomic system, the real and imaginary parts of the refractive index are described by $n_R=\chi'/2$ and $n_I=\chi''/2$, respectively \cite{Zhang2016,Zhang2018b}. As a result, a photonic graphene lattice governed by $n_R(x,y)$ is effectively “written” inside the medium. The refractive index variation and absorption spectrum are polarization-dependent. As presented in Fig.~3 of Ref. \cite{Zhu1999}, for the co-polarized configuration of the coupling and probing beams, the absorption of the probe beam around zero detuning is greatly reduced by the depletion of the ground state with optical pumping, whilst hardly any EIT effect occurs. For the cross-polarized configuration, there is clearly the EIT effect around zero detuning, while the absorption of the probe beam is also relatively strong. As a result, around zero detuning, $\chi'$ and $\chi''$ are approximately 5 and 20 times smaller for the co-polarized component (depleted, $y$) than for the cross-polarized one (EIT, $x$). The characteristic scales are $\chi'_x\sim 10^{-3}$, $\chi''_x\sim 10^{-4}$ and $\chi'_y\sim 2\times 10^{-4}$, $\chi''_y\sim 5\times 10^{-6}$ (for a $y$-polarized pump, i.e. the coupling beam).

The A- and B-type sublattices of the optically induced two-dimensional photonic graphene are marked in Fig.~\ref{fig1}(e). By selectively covering only the A or the B sublattice with the periodic probe field as in Fig.~\ref{fig1}(c), the $K$ or $K'$ valley in the momentum space can be effectively excited, which, due to the beam conversion between the sublattices, introduces an orbital angular momentum (OAM) to one of the output probe beams \cite{Song2015,Zhang2019}, confirmed by a fork-like feature in the interference pattern with a Gaussian reference beam (from the same laser as probe beams). Both the transmitted probe beam and phase (interference pattern) are monitored by a charge coupled device camera (CCD). To investigate the influence of the polarization of the probe beams on the OAM creation from the valley pseudospin, a quarter-wave plate is applied on the path of the probe beams before entering the atomic vapor cell, allowing varying the probe polarization. It should be noted that regardless of the probe beam polarization, the CCD always detects a single linear polarization component ($x$) defined by the PBS before it.

\section{Results and Discussion}
We are now considering polarized light described by a two component wave-function propagating in the photonic graphene lattice. %The graphene eigenstates are described a two component wave-function associated with the amplitude on the A and B-sites respectively. So in total we are describing a four components system. 
In general, solving 2D paraxial equations is much more efficient than solving the full system of Maxwell's equations in 3D. We begin our analysis with the comparison of the results of the two different numerical approaches for the simple case of a Gaussian probe of a width $w$ exciting the vicinity of the Dirac point of graphene, and neglecting the photonic SOC. In such a case the evolution of the beam is well described by the 2D Dirac equation describing the coupling between $A$ and $B$ sites of the honeycomb lattice. 
This approximation remains valid even including SOC in both excitation and detection are carried out in the same linear polarization. The simulated probe is exciting only $A$-sites of the graphene lattice, and its conversion to the $B$-sites is accompanied by the change of angular momentum according to $l_B=l_A-1$. This conversion is due to the non-zero Berry curvature of the graphene bands in the vicinity of the Dirac point \cite{Chang2008}, as shown by previous studies \cite{Song2015,Zhang2019}. This behavior can be obtained both with the 3D FEM and 2D paraxial equations, as shown in Fig.~\ref{fig2}(a,c), and, naturally, with the Dirac equation (which is an extra approximation with respect to the paraxial equation). 
%In these conditions, neither the absorption nor the spin-orbit coupling play any role, because the detected polarization component $x$ remains dominant. 
The interference patterns in panels (a,c) show the forklike dislocation which is a signature of a non-zero final angular momentum: $l_B=-1$ (initially, $l_A=0$). We note that the $A\to B$ conversion period $T=w/c$ depends on $\chi$ via the effective speed of light in the Dirac equation $c$.

\begin{figure}[tbp]
\centering
\fbox{\includegraphics[width=\linewidth]{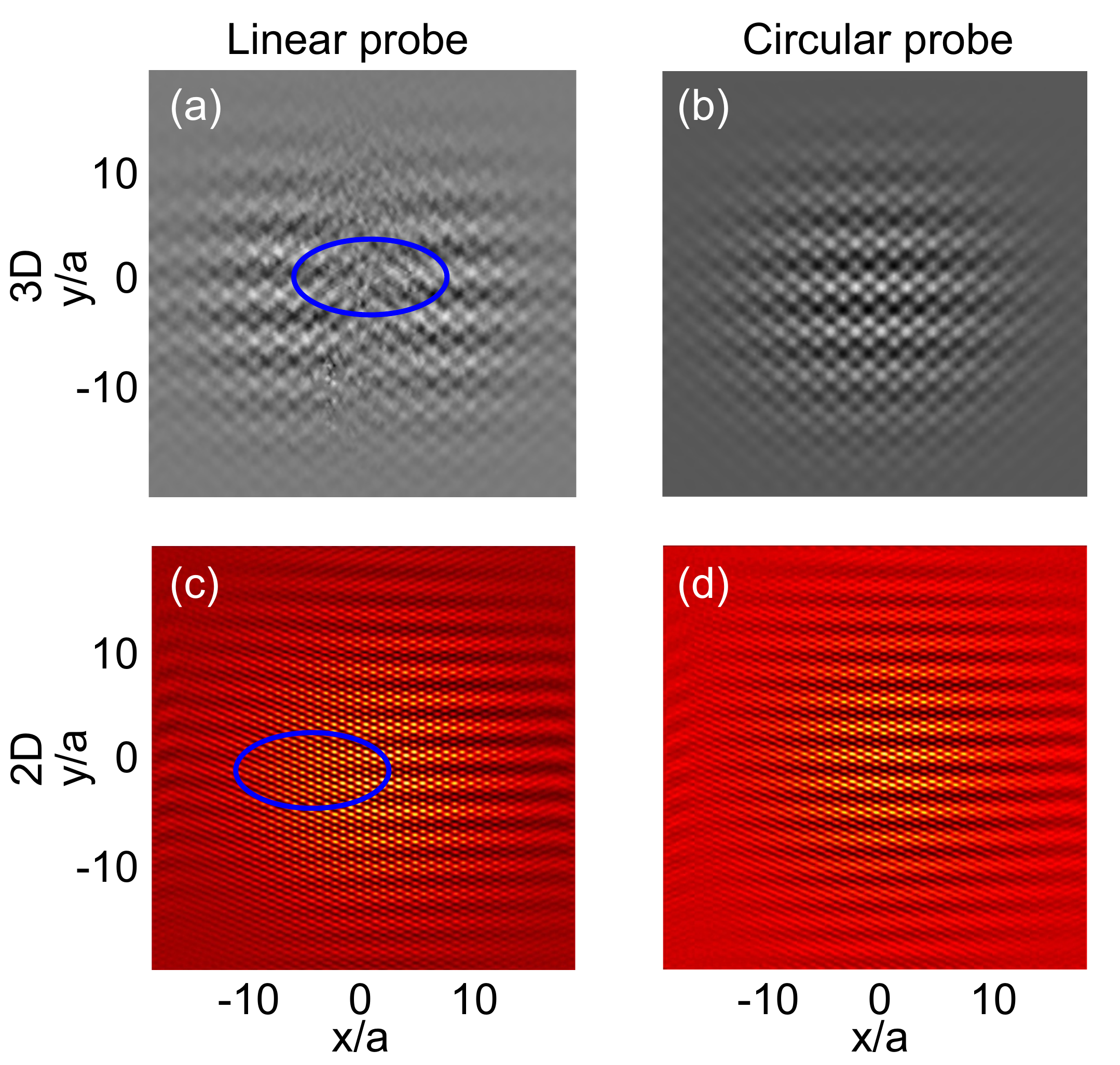}}
\caption{Theoretical simulation of interference patterns between the reference and probe beams after photonic graphene: 3D (a,b) and 2D (c,d) models: linear (a,c) and circular (b,d) probe. The dislocations indicating optical vortices are present only for linear probe.}
\label{fig2}
\end{figure}

The situation changes when the excitation contains both $E_x$ and $E_y$ (circular pumping) and the SOC starts to play a role. Figure~\ref{fig2}(b,d) shows interference patterns between the transmitted amplitude in the $x$-polarization and a reference beam. It does not show any dislocations in the fringe patterns. The absence of angular momentum conversion is a joint result of 3 different effects. First, the $y$-polarized component exhibits a slower $A\to B$ conversion because of a smaller real susceptibility $\chi'_y\approx 0.2\chi'_x$. So $E_y$ does not change angular momentum during the propagation time in the cell. The second effect is the smaller decay of the y-component,  because of a smaller imaginary susceptibility $\chi''_y\approx0.05\chi''_x$. As a result, the $E_y$ component quite rapidly becomes dominant over $E_x$ and most of the $x$-polarized light is induced by the $E_y$ to $E_x$ conversion by the  SOC. Ultimately, the $E_x$ generated by the SOC is not affected by the angular momentum conversion because the corresponding wave function is not anymore close to the Dirac point but lies in an upper band of graphene as explained below. As a result of these combined processes, the angular momentum $L=0$ measured in $E_x$ in Fig.~\ref{fig2}(b,d) is that of $l_A$ injected in $E_y$ and we can conclude that the angular momentum of the light after the cell is controlled by the incoming polarization.

Let us first make a simple estimate proving that the contribution of the polarization conversion can indeed become dominant with respect to the rapidly decaying original signal. The relevant terms in the paraxial equation for the description of the intensity contributions to the detected polarization $E_x$ become 
\begin{equation}
    i\frac{\partial E_x}{\partial z}=-\frac{i k_0\chi''_x}{2} E_x
+\frac{1}{2k_0}\frac{\partial \chi'_y}{\partial y}\frac{\partial E_y}{\partial x}E_y
\end{equation}
The first term provides an exponential decay with a characteristic rate $k_0\chi''_x\approx 10^7\times 10^{-4}=10^3$~m$^{-1}$, while the second one provides a linear growth (assuming $E_y=const$ as compared with $E_x$). To estimate its rate, we use the experimental parameters $k_x/k_0\approx\theta\approx 10^{-2}$ and $\partial \chi'_y/\partial x\approx 10^{-4}/25\times10^{-6}= 4$, which gives a rate of the order of $10^{-2}$. Comparing an exponential decay $\exp(-10^3z)$ with a linear growth $10^{-2}z$ for identical initial intensity (circular pumping), we see that the two contributions become equal within a propagation distance of $1$~cm. We can therefore conclude that for a 5~cm vapor cell the contribution of the SOC-induced polarization conversion can indeed be important and even dominant. A more detailed discussion of the evolution of the intensities based on rate equations is given in the Supplemental materials.

The other crucial effect is the absence of angular momentum conversion for the $E_x$ polarization induced by SOC. This can be understood by using arguments based on tight-binding description (while the development of a complete tight-binding model accounting for the specific SOC is beyond the scope of the present work). The SOC terms contain the first-order derivatives of the wave function. It thus inverts the mode symmetry, coupling the $s$ (symmetric) and $p$ (anti-symmetric) orbitals of each minimum of the effective lattice potential. Indeed, in Fig.~\ref{fig0} a single maximum (Gaussian, $s$-orbital) was converted into an anti-symmetric double-maximum structure ($p$-orbital). The converted polarization therefore appears at the same valley, but in the  $p$ band. The shape of the wave-function suggests that conversion occurs toward the flat $p$ band of the honeycomb potential\cite{Jacqmin2014} where further evolution is completely blocked (see the Supplemental materials for more details). This is why the converted component maintains its original angular momentum. The observed signal is the superposition of the original component $E_x$ (changing the angular momentum) and the signal converted from $E_y$ (keeping the original angular momentum). The final result depends on the relative intensities of the two components, and varying their initial ratio  (which is the circular polarization degree) allows to observe the vortex leaving the beam center step by step, when the circular polarization is increased (see below for the experimental results), similar to the spatial dynamics observed in Ref.~\cite{Zhang2019}, but in opposite direction.

\begin{figure}[tbp]
\centering
\fbox{\includegraphics[width=\linewidth]{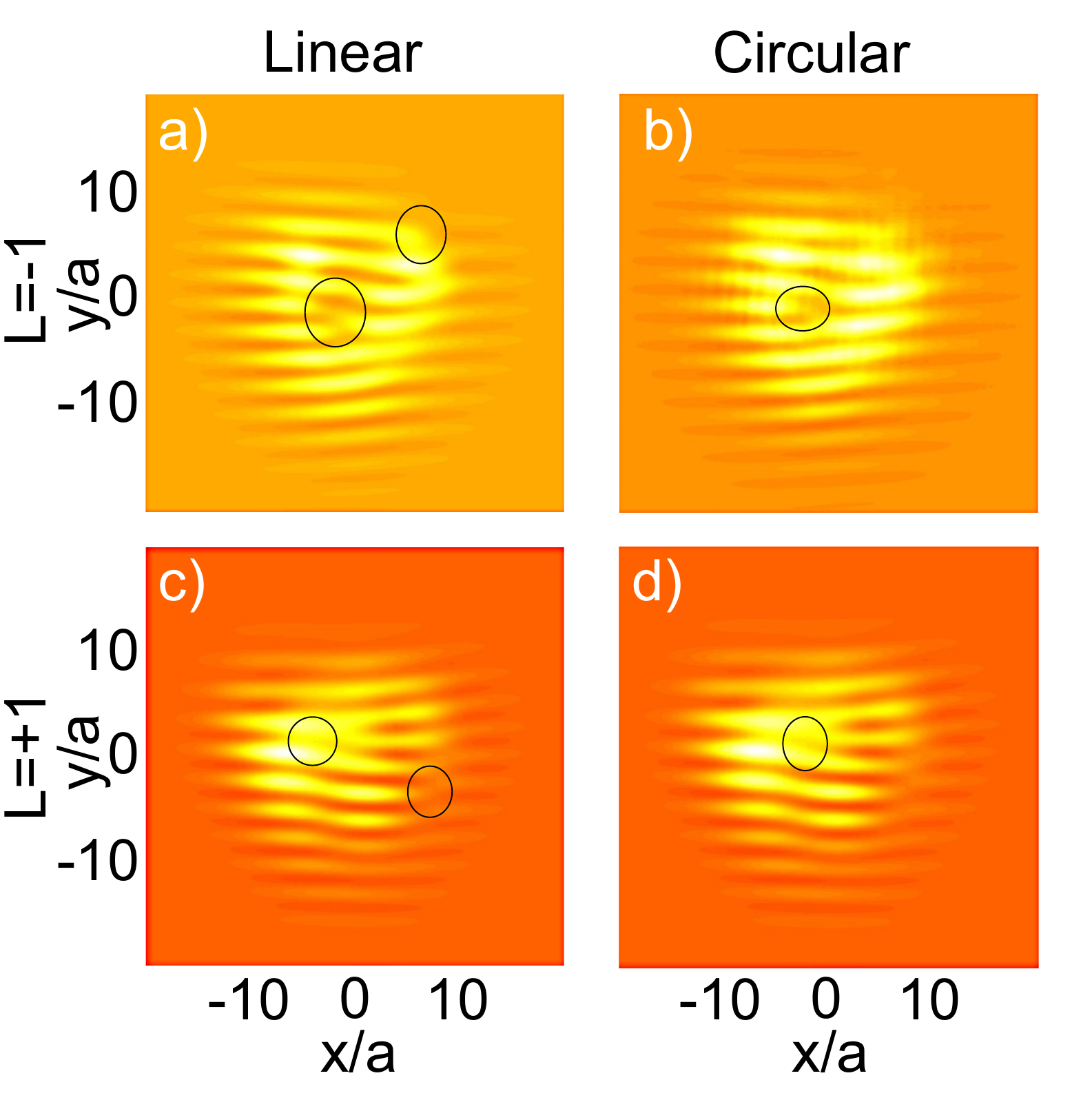}}
\caption{Theoretical simulations of the interference patterns of separated single-valley beams with a reference beam (2D paraxial equation). Different initial angular momentum: $L=+1$ (a,b) and $L=-1$ (c,d); different initial polarization: circular (a,c) and linear (b,d).}
\label{fig3}
\end{figure}

The comparison of the 3D and 2D results demonstrates the correctness of the developed corrected paraxial equations. In what follows we are are going to use only the 2D model and to consider a probe with non-zero initial angular momentum. Figure~\ref{fig3} shows the results of numerical simulations with 2D coupled paraxial equations for the case of Gauss-Laguerre probe envelope with $L=-1$ (a,b) and $L=+1$ (c,d) with linear (a,c) and circular (b,d) probe polarization. For $L=-1$ and linear excitation, the beam conversion introduces an extra vortex into the pattern, giving $L_{out}=-2$. For $L=+1$, the sign of the extra vortex is opposite to the initial angular momentum, and $L_{out}=0$. This extra vortex disappears in both cases if the initial polarization is circular, because the output beam is dominated by the $E_y$ converted to $E_x$. A crucial advantage of using a 2D model here is that it allows a more extended treatment of the output field distributions, in particular, valley separation and interference analysis (based on Fourier transform and shifting in the reciprocal space), similar to the beam separation after the vapor cell in the experiment \cite{Zhang2019}.

\begin{figure}[tbp]
\centering
\fbox{\includegraphics[width=\linewidth]{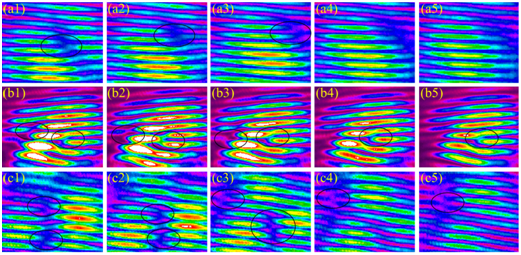}}
\caption{Experimental interference images for different GL beams: a) $L=0$; b) $L=-1$; c) $L=+1$. Numbers correspond to circular polarization degree: from $\rho_c=0$ (linear probe, a/b/c1) to $\rho_c=1$ (circular probe, a/b/c5). In each case, the change of polarization induces the decrease of the output beam angular momentum by 1.}
\label{fig4}
\end{figure}

Finally, in Fig.~\ref{fig4} we present the results of experimental measurements carried out in the configuration described in the previous section, for 3 values of the angular momentum of the inbound Gauss-Laguerre probe: a) $L=0$ (as in Fig.~\ref{fig2}, b) $L=-1$, c) $L=+1$ (as in Fig.~\ref{fig3}). The false color scale represents the intensity of the interference pattern, measured at the screen after the vapor cell in the $x$ polarization. The interference occurs between one of the two probe beams and a reference beam. The numbered panels represent the increase of the circular polarization degree from $\rho_c=0$ (a1,b1,c1) to $\rho_c=1$ (a5,b5,c5). The experimental observations correspond to the theoretical images shown above: in each case, the extra vortex brought into the beam by the interaction with graphene lattice disappears when $\rho_c$ is increased, because signal is dominated by the converted component.

In all three cases we observe that the SOC allows controlling the angular momentum of the output beam via both the polarization and momentum  of the input beam. Devices with an optically-controllable optical angular momentum  are useful for practical applications, but creating them is a challenging task \cite{Zambon2019}. Our experiment provides a solution of this problem, while demonstrating the fundamental importance of SOC in the paraxial approximation, which has been neglected so far. In the future works, we will develop the tight-binding description of this original type of SOC.

To conclude, we introduce a first-order correction to paraxial equations due to the SOC linked with the conservation of the polarization plane. We demonstrate experimental evidence of the importance of this SOC in photonic graphene implemented via EIT in atomic vapors. The angular momentum of the output beam can be controlled by the circular polarization of the probe.

\section*{Funding Information}
This work was supported by National Key R\&D Program of China (2018YFA0307500, 2017YFA0303703), National Natural Science Foundation of China (11804267,61975159), and Fundamental Research Funds for the Central Universities (xjj2017059). We acknowledge the support of the project "Quantum Fluids of Light"  (ANR-16-CE30-0021), of the ANR Labex Ganex (ANR-11-LABX-0014), and of the ANR program "Investissements d'Avenir" through the IDEX-ISITE initiative 16-IDEX-0001 (CAP 20-25). 

%\section*{Acknowledgments}
%We thank X for discussions.

\medskip

\noindent\textbf{Disclosures.} The authors declare no conflicts of interest.

\section*{Supplemental Documents}

\bigskip \noindent See \href{link}{Supplement 1} for supporting content.

% Bibliography
\bibliography{biblio}

\section{Supplemental Materials}

In this supplemental material, we present additional results concerning the polarization conversion rate, the mixing of the $s$ and $p$ bands by the spin-orbit coupling, and the sensitivity of the qualitative results to the input parameters.

\subsection{Rate equations for polarization conversion}

In the main text of the manuscript, we discuss the relative contribution of the original and converted polarization components to the detected intensity. Here, we use the rate equations equivalent to the paraxial equations in order to describe the behavior of the intensity and to confirm that ultimately all measured intensity originates from converted polarization (under circular pumping).

The rate equations for the intensities of the $x$ and $y$ polarizations $I_x(t)$, $I_y(t)$ can be written as:
\begin{eqnarray}
\frac{dI_x}{dt}&=&-I_x\Gamma_x+W\left(I_y-I_x\right)\\
\frac{dI_y}{dt}&=&-I_y\Gamma_y+W\left(I_x-I_y\right)
\end{eqnarray}
This system of differential equations can be solved analytically, but in order to have less cumbersome expressions we shall apply several approximations. First, since $\Gamma_x\gg\Gamma_y$, we can neglect $\Gamma_y$. Second approximation is based on the fact that the polarization conversion rate $x\to y$ is much smaller than the decay of the $x$ polarization: $W\ll\Gamma_x$ (see the main text for the discussion of the numerical values). This allows to neglect the second-order term using $W/\Gamma_x\ll 1$. Keeping only the first order terms in $W/\Gamma_x$, the solution for the detected component $I_x(t)$ can be written as:
\begin{equation}
    I_x(t)=I_0\left(e^{-\Gamma_x t}e^{-Wt}\left(1-\frac{W}{\Gamma_x}\right)+e^{-Wt}\frac{W}{\Gamma_x}\right)
    \label{solxy}
\end{equation}
The physical meaning of this solution is clear: the first term corresponds to the original polarization, which decays even faster than just $\Gamma_x$ because of the additional losses due to the conversion to the other component. The second term corresponds to the intensity generated from the second component, and its decay rate is determined by $W$.

To get the information on the fraction of the intensity $I_x$ coming from the conversion from the other component $I_y$, we can compare the solution~\eqref{solxy} to the behavior of an isolated polarization component $I_i$ with the same decay rate $\Gamma_x$, given by $I_i(t)=I_0\exp(-\Gamma_x t)$. The expression for this fraction reads
\begin{equation}
    \frac{\Delta I}{I}=\frac{e^{-\Gamma_x t}e^{-Wt}\left(1-\frac{W}{\Gamma_x}\right)+e^{-Wt}\frac{W}{\Gamma_x}-e^{-\Gamma_x t}}{e^{-\Gamma_x t}e^{-Wt}\left(1-\frac{W}{\Gamma_x}\right)+e^{-Wt}\frac{W}{\Gamma_x}}
\end{equation}
This expression can be simplified again by using $W\ll\Gamma_x$. At small times, the series expansion of the exponents gives
\begin{equation}
    \frac{\Delta I}{I}\approx W t
\end{equation}
and at longer times
\begin{equation}
    \frac{\Delta I}{I}\to 1
\end{equation}
We conclude that in the regime of interest, the fraction of the converted component grows linearly with time, with a rate directly determined by the efficiency of the polarization conversion.

\subsection{Band mixing by spin-orbit coupling}

\begin{figure}[tbp]
\centering
\fbox{\includegraphics[width=\linewidth]{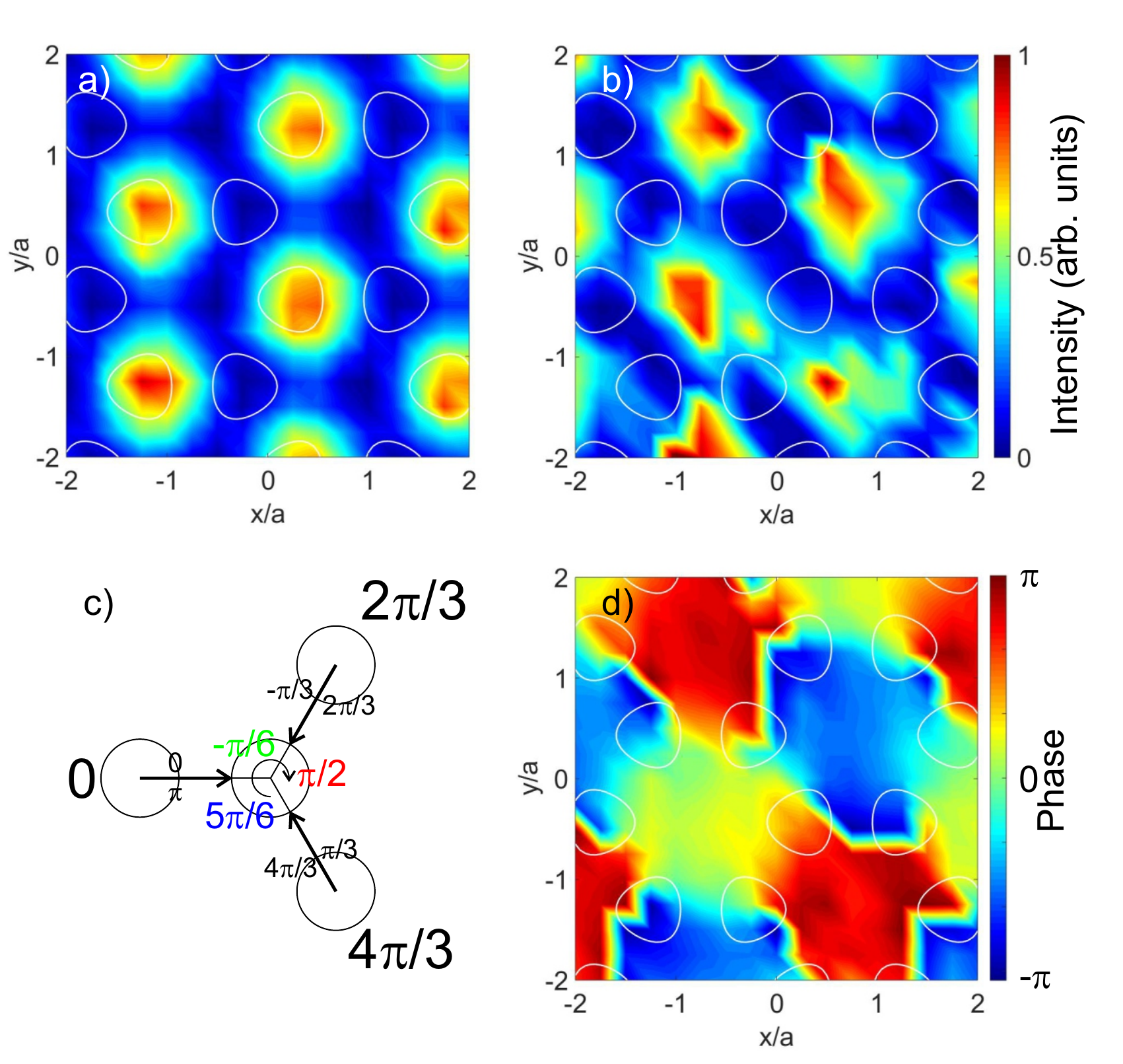}}
\caption{a) Initial intensity distribution of $E_y^2$; b) Final intensity distribution of $E_x^2$; c) Scheme of the phases of the converted polarization; d) Final phase distribution of $E_x$. }
\label{fig1s}
\end{figure}

In this section, we explain the microscopic mechanism of the mixing of the $s$ and $p$ bands of the photonic graphene by the spin-orbit coupling appearing due to the susceptibility gradient.

As an illustration, we perform a 3D FEM simulation of the beam evolution in the photonic graphene lattice using COMSOL, as in Fig.~3(c) of the main text. However, here we simulate the excitation of the system with orthogonal linear polarization $E_y$, in order to elucidate clearly how it evolves in the lattice. Moreover, we simulate a 3-beam excitation instead of a 2-beam one, in order to be able to make conclusions on the band mixing by studying the intensity and phase distribution.

Figure~\ref{fig1s}(a) shows the distribution of intensity $E_y^2$ at the input of the vapor cell. The susceptibility profile responsible for the effective potential of the honeycomb lattice is shown with white contour lines. The maxima of intensity appear at the positions of the sites of a single type only (the $A$-sites), as expected for a state of the s-band. Figure~\ref{fig1s}(b) shows the intensity in the x-polarisation at the output of the cell. The maxima are now located at the center of each unit cell, corresponding to a state of the $p$-band, as expected from the symmetry of the spin-orbit coupling. To explain the phase distribution in the final state, we present a scheme showing the microscopic mechanism of its formation in Fig.~\ref{fig1s}(c). The wavefunction of the $K$ point of the $s$-band is initially distributed over the $A$-sites with a phase forming a vortex around the center of the unit cell (phase values shown at each site). Because of the effective potential due to the susceptibility gradient, the intensity expands in the directions shown by the arrows, but the phase of the generated wave is opposite at each side of each arrow (anti-symmetric function). The final values of the phase in each sector are shown with colored numbers. They correspond to a vortex around the empty $B$-site. This is confirmed by the phase $\arg E_x$ extracted from the numerical simulations, shown in Fig.~\ref{fig1s}(d). This Bloch function corresponds to the same valley as the one of the $s$-band, but now it belongs to the $p$-band, because the particles are located at the barriers of the potential, and not at its minima. The shape of the orbital corresponds to the frustrated states from the flat $p$-band of photonic graphene studied experimentally and theoretically in Ref.~\cite{Jacqmin2014}.

We note that the vortices forming in the Bloch function at each unit cell should not be confused with the vortex forming in the envelope function! The spatial separation of the beams forming the Bloch function after the vapor cell allows to get access to the envelope function. In simulations, this is done using Fourier transform and $k$-space shifting.

We have tested this configuration (with only $E_y$ excitation) experimentally as well, and these experiments confirm the blocking of the formation of the vortex in the envelope function, which we explain by the coupling with the flat $p$-band.

\subsection{Sensitivity to parameters}

The susceptibility of the atomic vapors under the effect of optical pumping is not known with a very high accuracy. This especially concerns its polarization dependence. While the qualitative ratio $\chi''/\chi'\sim  0.1$, $\chi_y/\chi_x\sim 0.1$ is generally accepted, the exact ratios are difficult to determine and depend on the detuning of a particular experiment. The results shown in the main text correspond to $\chi'_x=5\chi'_y$ and $\chi''_x=20\chi''_y$. In this section, we show that qualitatively the same results are obtained if these ratios are taken equal to $10$ and $10$, respectively.

\begin{figure}[tbp]
\centering
\fbox{\includegraphics[width=\linewidth]{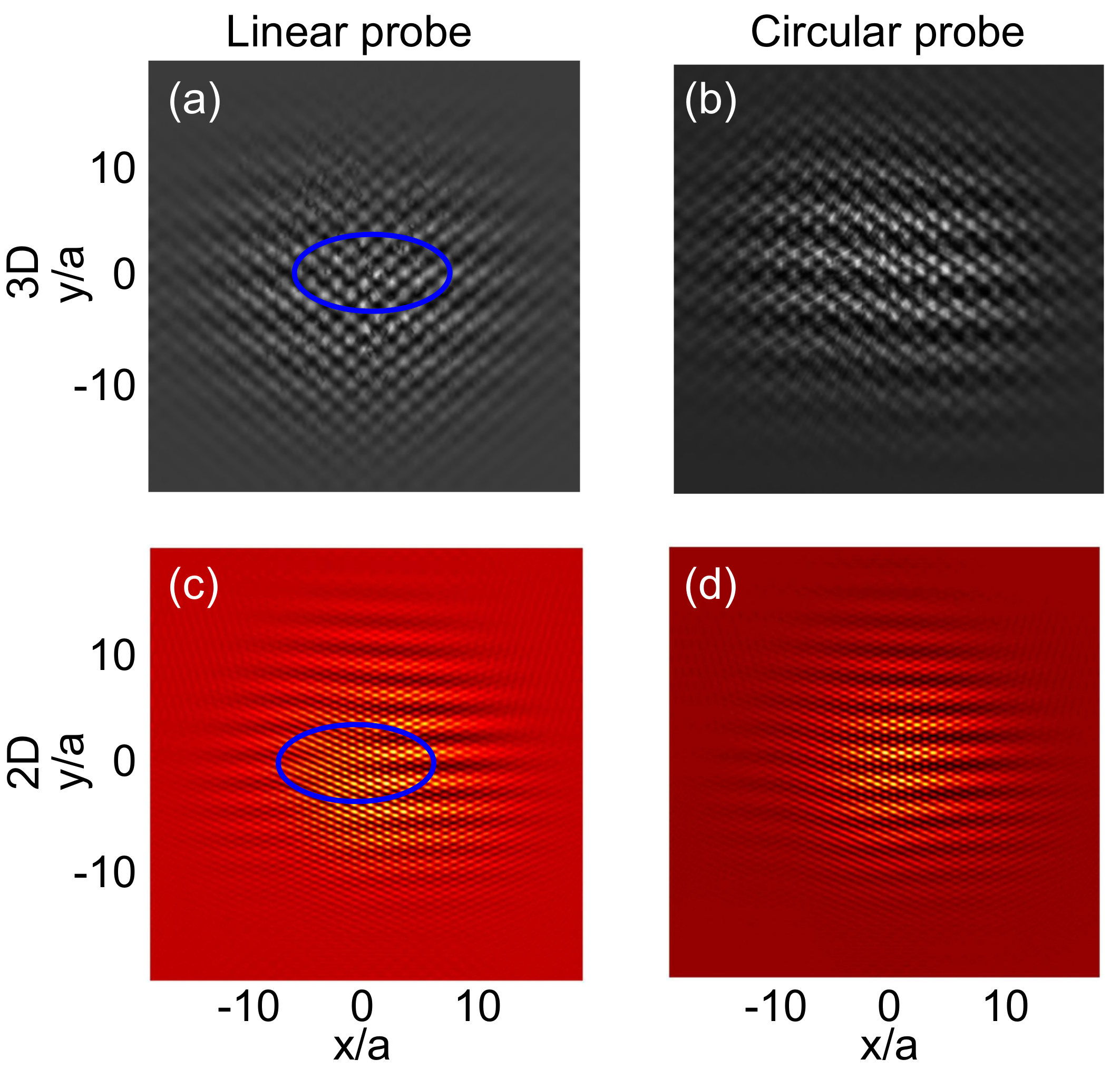}}
\caption{Theoretical simulation of interference after photonic graphene: 3D (a,b) and 2D (c,d) models: linear (a,c) and circular (b,d) probe. The dislocations indicating optical vortices are present only for linear probe.}
\label{fig2}
\end{figure}

Figure~\ref{fig2} shows the comparison of the 3D (FEM beam envelope, (a) and (b)) and 2D (paraxial, (c) and (d)) models for a Gaussian incident beam ($L=0$) with linear ((a) and (c)) and circular ((b) and (d)) polarization. While the distribution of intensity is slightly modified with respect to the main text, the qualitative behavior is conserved: the generated vortex observed in the envelope function for the linear probe disappears for the circular probe.

\begin{figure}[tbp]
\centering
\fbox{\includegraphics[width=\linewidth]{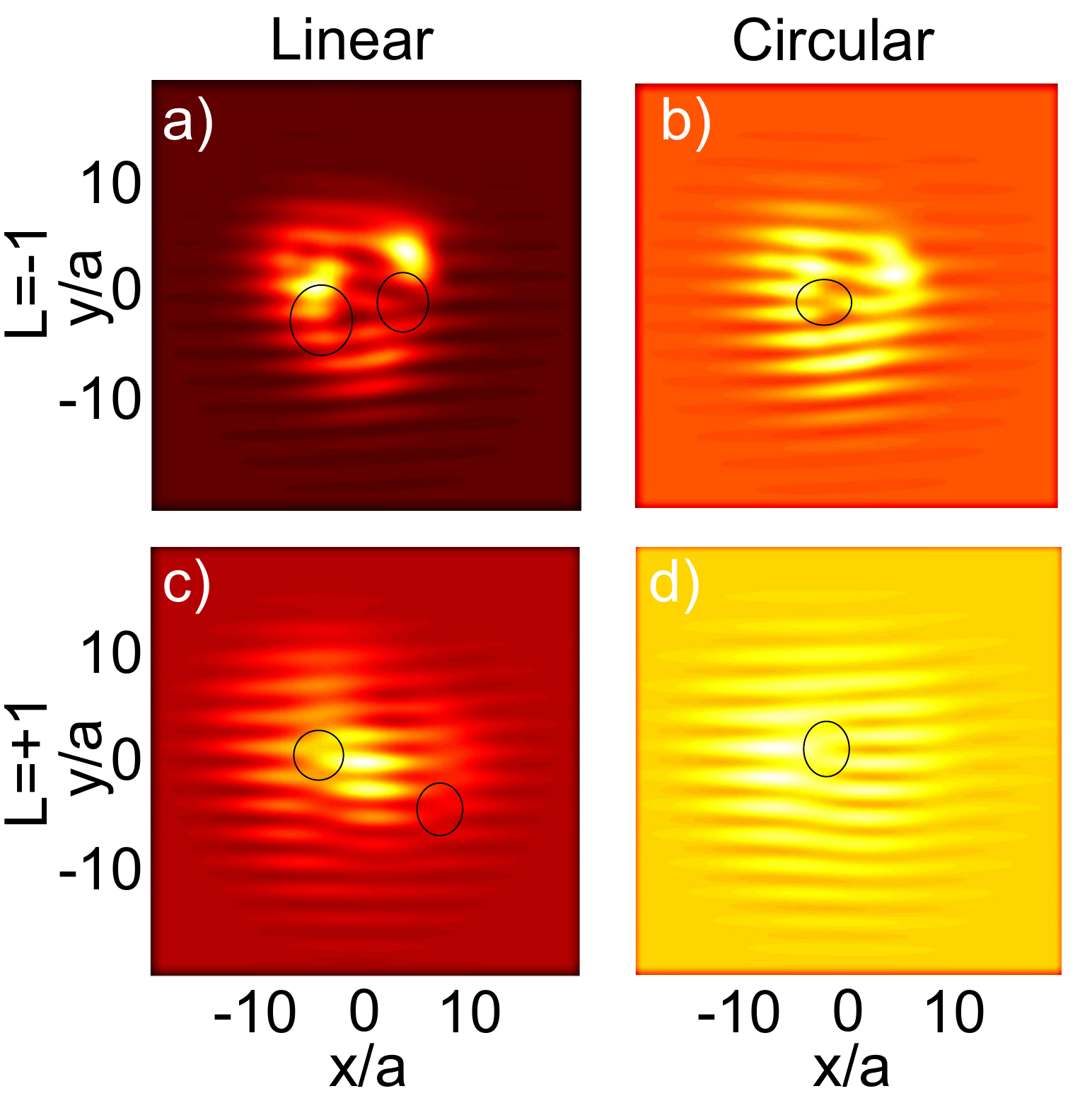}}
\caption{Theoretical simulations of the interference patterns of separated single-valley beams (2D paraxial equation). Different initial angular momentum: $L=+1$ (a,b) and $L=-1$ (c,d); different initial polarization: circular (a,c) and linear (b,d).}
\label{fig3}
\end{figure}

Figure~\ref{fig3} shows the results obtained by numerical simulations based on the 2D paraxial equation for the Gauss-Laguerre envelope beams $L=\pm 1$. In both cases, the angular momentum changes by 1 under linear pumping (negligible SOC), while this change does not occur under circular pumping (SOC-dominated regime). The behavior is therefore the same as in the main text.

% Full bibliography will be added automatically on a new page for Optics Letters submissions. This command is ignored for journal article submissions.
% Note that this extra page will not count against page length.
%\bibliographyfullrefs{biblio}

%Manual citation list
%\begin{thebibliography}{1}
%\bibitem{Zhang:14}
%Y.~Zhang, S.~Qiao, L.~Sun, Q.~W. Shi, W.~Huang, %L.~Li, and Z.~Yang,
 % \enquote{Photoinduced active terahertz metamaterials with nanostructured
  %vanadium dioxide film deposited by sol-gel method,} Opt. Express \textbf{22},
  %11070--11078 (2014).
%\end{thebibliography}

\end{document}